\documentclass[conference]{IEEEtran}
\usepackage{graphicx} 
\usepackage{cite}
\usepackage{float}
\usepackage{url}
\usepackage{siunitx}
\usepackage{comment}
\usepackage{amsmath}
\usepackage{booktabs}
\usepackage[section]{placeins}
\title{Quantitative Analysis of UAV Intrusion Mitigation for Border Security in 5G with LEO Backhaul Impairments}

\author{
\IEEEauthorblockN{Rajendra Upadhyay, Al Nahian Bin Emran, Rajendra Paudyal, Lisa Donnan, Duminda Wijesekera}
\IEEEauthorblockA{
\text{Mason Innovation Labs, 
George Mason University, Fairfax, VA, 22030, USA}\\
\text{\{rupadhya $\mid$ abinemra $\mid$ rpaudyal $\mid$ ldonnan $\mid$ dwijesek@\}gmu.edu}
}
}
\def\BibTeX{{\rm B\kern-.05em{\sc i\kern-.025em b}\kern-.08em
    T\kern-.1667em\lower.7ex\hbox{E}\kern-.125emX}}

\makeatletter
\newcommand{\linebreakand}{%
  \end{@IEEEauthorhalign}
  \hfill\mbox{}\par
  \mbox{}\hfill\begin{@IEEEauthorhalign}
}
\makeatother

\begin{document}

\maketitle
\begin{abstract}
Uncooperative unmanned aerial vehicles (UAVs) pose emerging threats to critical infrastructure and border protection by operating as rogue user equipment (UE) within cellular networks, consuming resources, creating interference, and potentially violating restricted airspaces. This paper presents minimal features of the operating space, yet an end-to-end simulation framework to analyze detect-to-mitigate latency of such intrusions in a hybrid terrestrial-non-terrestrial (LEO satellite) 5G system. The system model includes terrestrial gNBs, satellite backhaul (with stochastic outages), and a detection logic (triggered by handover instability and signal quality variance). A lockdown mechanism is invoked upon detection, with optional local fallback to cap mitigation delays. Monte Carlo sweeps across UAV altitudes, speeds, and satellite outage rates yield several insights. First, satellite backhaul outages can cause arbitrarily long mitigation delays, yet, to meet fallback deadlines, they need to be effectively bounded. Second, while handover instability was hypothesized, our results show that extra handovers have a negligible effect within the range of parameters we considered. The main benefit of resilience from fallback comes from the delay in limiting mitigation. Third, patrol UEs experience negligible collateral impact, with
handover rates close to terrestrial baselines. Stress scenarios further highlight that fallback is indispensable in preventing extreme control-plane and physical security vulnerabilities: Without fallback, prolonged outages in the satellite backhaul delay lockdown commands, allowing rogue UAVs to linger inside restricted corridors for several seconds longer. These results underscore the importance of complementing non-terrestrial links with local control to ensure robust and timely response against
uncooperative UAV intrusions.
\end{abstract}


\section{Introduction}
\label{sec:introduction}

UAVs and drones are increasingly accessible and capable in a variety of applications, from logistics, smart agriculture, to emergency response applications. However, their potential misuse in sensitive
areas such as borders, airports, and critical infrastructure protection \cite{emran2025tpm} poses significant security challenges. In particular, uncooperative UAVs/drones that are not registered, are not whitelisted can actively enter restricted airspace and interfere with legitimate wireless services for hostile purposes.

Mobile networks, especially emerging 5G and 6G systems, provide an opportunity and a challenge to address this threat. Opportunistically, the 5G/6G system's ability to densely deploy base stations, direct beamform on demand, and integrate terrestrial and NTNs to offer a wealth of detection capabilities and flexible pathways for mitigation. On the challenge side, outage-prone satellite backhaul, heterogeneous mobility problems~\cite{rahman2023pacman}, and variations in handover dynamics complicate timely and reliable detection of intruders.

\subsection{Problem Statement}  
\label{sec:problem}

The central question addressed in this work is detecting and mitigating uncooperative UAVs entering a protected corridor while minimizing collateral impact on legitimate ground users. 
Uncooperative UAVs can take several forms. Small commercial drones may be used to smuggle contraband across borders, while modified UAVs can act as airborne surveillance platforms or even carry hazardous payloads. Such devices are not registered with the network and do not appear on any authorized whitelist, yet still interact with the cellular radio environment. Because aerial users at altitude experience strong line-of-sight connectivity to multiple base stations, they generate mobility signatures that differ markedly from patrol UEs, such as frequent handovers, higher neighbor cell counts, and large Reference Signal Received Power (RSRP) variance. These anomalies form the basis for the detection in our model.

Once detected, the mitigation mechanism considered here is a \emph{lockdown} procedure: the UAV is forcibly attached to the gNB that serves it, preventing further handovers or continued exploitation of the network. Under terrestrial backhaul conditions, such commands can be executed immediately. However, when the serving gNB is based on LEO satellite backhaul, mitigation commands must traverse the satellite link. This introduces additional latency and the risk of several seconds of outages during satellite handovers or gateway transitions. In such cases, mitigation is delayed, allowing the intruding UAV to linger within the protected corridor and potentially violate no-fly zones before lockdown can be enforced.

This interplay between detection, mitigation, and backhaul reliability motivates our study to evaluate how LEO outages impact the detect-to-mitigate loop and to assess if a local fallback deadline can bound these delays and ensure a timely response against UAV intrusions.

\subsection{Summary of Contributions.}
\label{ssec:ContributionSummary}

This paper develops and evaluates a minimal end-to-end simulation framework tailored for this scenario. The key contributions of the work are as follows.
\begin{itemize}
    \item We coded a minimal simulator in MATLAB with three gNBs, one or all of which can have LEO backhaul. It is intentionally kept simple so that we can isolate the effect of outages on mitigation delays.
    \item Monte Carlo simulation takes advantage of the speed, altitude, and network scenarios of UAVs, which can analyze situations that stress the detection and mitigation techniques, which we call stress cases.
    \item Empirical evaluation of key performance indicators (KPIs) such as detection-to-mitigation delay, excess handovers, and collateral consequences of patrol UE instability.
    \item Demonstrate that a local fallback deadline (e.g., 2~s) effectively bounds mitigation delay and prevents catastrophic failures during long satellite outages.
\end{itemize}

By focusing on a minimal but interpretable set-up, this study provides quantitative evidence for the value of hybrid terrestrial and non-terrestrial architectures in border protection. Our current results are intended not as a
final system design, but as an initial step towards creating a more realistic multi-UAV and multi-cell evaluation, guiding both academic research and early standardization efforts.

\section{Background}
\label{sec:background}

Unmanned Aerial Vehicles (UAVs, or \emph{drones}) have emerged as both a valuable tool and a security concern in border protection. Border security agencies deploy UAVs for surveillance and patrol, leveraging their aerial perspective for real-time monitoring of remote or difficult terrain~\cite{koslowski_schulzke_drones_along_borders}. However, adversaries exploit the same technology, such as criminal organizations using small, uncooperative drones to smuggle contraband and conduct clandestine surveillance of patrol movements. There are recent reports of hundreds of unauthorized drone incursions per month on the US-Mexican border, where some have even been modified as kamikaze devices carrying explosives~\cite{airsight_border_security_drones}. This dual-use scenario underscores the urgent need for robust intrusion detection and mitigation mechanisms tailored to unauthorized UAVs in border airspace.

Fifth-generation (5G) cellular networks offer a promising backbone for UAV communication and control, including beyond the visual line of sight operations. Compared to unlicensed Wi-Fi links traditionally used by drones, 5G provides broader coverage, regulated spectrum, and built-in security, reliability, and fallback features. These characteristics make 5G an attractive alternative for drone command and control and data streaming that is capable of supporting the high throughput and low latency that some UAV applications require~\cite{handoverchallenges}. The 3GPP began improving cellular standards to better accommodate aerial users in LTE Release~15 and beyond ~\cite{3gpp_nr_uav_2023}. Recent 5G specifications introduce UAV-specific features, for example, height-aware measurement events and flight-path reports, to ensure reliable connectivity for drones aloft~\cite{3gpp_nr_uav_2023}. Networks can even request the planned route of a drone to anticipate its cell transitions~\cite{3gpp_nr_uav_2023}. In addition, mechanisms for UAV identification and authorization have been defined. Only drones with proper network subscriptions must be served, and a broadcast UAV ID (over sidelink PC5) is envisioned to help authorities detect any UAV in a given area~\cite{3gpp_nr_uav_2023}.These advancements facilitate legitimate UAV operations. In contrast, an uncooperative UAV (one that lacks authorization and may not transmit any ID) can still enter 5G coverage and pose a threat. Such rogue drones will not be granted normal network access, yet their presence may be inferred by the network through anomalous radio frequency (RF) signatures.

Crucially, UAVs behave very differently from typical ground user equipment (UE) in the radio access network (RAN). A drone flying at altitude often has a line of sight (LoS) to multiple base stations, observing strong signals from far away cells and causing greater interference and uplink noise than ground UEs~\cite{3gpp_nr_uav_2023}. One consequence is drastically increased handover activity and volatility in signal strength. Measurements in an LTE network showed that a drone at 150~m altitude experienced about five cell handovers per minute, compared to only one handover per minute for a ground vehicle at similar speed. This occurs because the number of base stations simultaneously visible to an aerial UE rises with height~\cite{handoverchallenges}, triggering frequent cell re-selections. Likewise, the received signal power (RSRP) of the drone can fluctuate rapidly as it transitions between overlapping coverage areas. Consequently, RAN-side metrics, such as handover frequency, neighbor cell count, RSRP variance, and others, naturally serve as indicators of an aerial intruder. In essence, an uncooperative UAV acts as an outlier in the mobility patterns of the cellular network. By monitoring mobile UEs that exhibit unusually frequent handovers or erratic signal strength (beyond what any patrol UE would), the network can passively detect the presence of a rogue drone. This approach effectively repurposes the 5G infrastructure as a sensor network for aerial threats, complementing traditional radar and optical drone detection systems.

Another challenge in using 5G for border UAV security lies in the backhaul network that connects remote cell sites. Border regions often rely on satellite links for backhaul due to the lack of fiber infrastructure. In particular, low-earth orbit (LEO) satellite constellations are increasingly being used to provide high-throughput wide-area backhaul for 5G in remote areas. LEO satellites offer much lower latency than traditional geostationary satellites on the order of 20--60~ms one-way making them viable for real-time applications. However, LEO-based backhaul comes with its own reliability issues. The latency, while low on average, is highly variable; Rapid changes in satellite links introduce significant jitter and occasional packet loss. Transient outages can occur due to satellite handovers, ground station switching, obstructions, and weather attenuation~\cite{cisco_leo_wp_2025}. In fact, even a brief outage of a few seconds on the control-plane link could be catastrophic in a security scenario that delays a critical command (such as an interdiction or lockdown where a suspicious UAV limited to a specific cell signal to handover-manage a rogue UAV) and allows the intruder to slip through. Thus, latency and outage resilience are paramount when applying 5G networks to time-sensitive UAV interception.

To mitigate these risks, 5G systems can employ fallback mechanisms that maintain operation during backhaul disruptions. An approach is \emph{multi-homed connectivity}. To explain by an example, we combine terrestrial and nonterrestrial links so that if the primary LEO backhaul is lost, an alternate path (such as a redundant satellite or a terrestrial microwave link) can temporarily carry the traffic. Another solution is to push critical control functionality to the network's edge. In a border protection context, this solution could allow the local base station (gNB) to autonomously execute certain mitigation actions if it cannot immediately communicate with the core or remote controller due to a backhaul outage. Such local fallback solutions ensure that time-critical responses (e.g. locking a suspicious UAV to a specific cell or broadcasting a takeover command) are not entirely dependent on an intact backhaul route. Similar concepts of local breakout and edge-resident control for ultra-reliable and low-latency operations have been demonstrated in industrial 5G networks \cite{paudyal2025gnb}. Fallback mechanisms thus act as a safety net, containing the threat until normal connectivity is restored. 

In summary, a combination of RAN-based detection and smart fallback control is essential for timely mitigation of uncooperative UAVs in border areas, especially under adverse network conditions introduced by LEO backhaul.

\section{Related Work}

\subsection{UAV Connectivity and Handover in 5G Networks}
\label{ssec:UAVconnectivity+handover}

A key strand of research examines how UAVs behave as aerial user equipment (UEs) in cellular networks, particularly with respect to handovers and mobility. 
Fakhreddine et.al. showed that altitude significantly increases the number of simultaneously visible base stations, causing more frequent handovers and interference compared to patrol UEs~\cite{handoverchallenges}. Muzaffar et al. extended this line of work with real-world experiments on a 5G-connected drone, reporting frequent handovers and variable flight performance compared to ground devices~\cite{muzaffar2020first}. 
Uddin et al. proposed the use of handover count statistics for the detection of mobility state, demonstrating that UAVs can be distinguished from patrol UEs using RAN KPIs alone~\cite{mobilitystatedetection}. 
Angjo et al. reviewed drone handover management approaches, including adaptive hysteresis, trajectory prediction, and ML-based methods~\cite{9319862}. On the standardization side, 3GPP has since introduced UAV-specific features in NR, such as height-aware measurements, UAV identification, and flight path reporting~\cite{3gpp_nr_uav_2023}. 
These works collectively establish that UAVs exhibit distinctive mobility patterns in cellular networks, which both complicate connectivity management and provide opportunities to estimate the behavior of these UAVs and anomaly detection.

\subsection{UAV Threats, Intrusion Detection, and Countermeasures in 5G}
\label{ssec:uavThreats+IDS}

Another strand of literature treats UAVs as threats to the network or as uncooperative intruders. 
Abdalla et al. analyzed UAV-assisted attacks against 5G networks and proposed prevention, detection, and recovery mechanisms, highlighting UAVs as both a platform for attacks and a detection target~\cite{abdalla2020uav}. 
In a complementary work, Abdalla and Marojevic reviewed emerging UAV communication standards, highlighting the need for regulatory compliance, UAV identification, and monitoring mechanisms to mitigate misuse~\cite{abdalla2020communications}. Targeted control-plane attacks such as energy-efficient control channel jamming have been explored in 5G contexts \cite{ashik2025reaperpulse}, underscoring the broader security exposure of wireless control mechanisms relevant to UAV-based threats.
Beyond the cellular view, several sensing modalities have been studied for non-cooperative UAV detection. 
Maksymiuk \emph{et al.} demonstrated a passive radar system using 5G NR downlink signals as illuminators for UAV detection~\cite{maksymiuk2022renyi}. 
Other IEEE studies explored the features of the RF spectrum for UAV detection and tracking~\cite{8741970}. 
Acoustic signatures have also been used for the same purpose. Busset et al. used advanced acoustic cameras to detect and track small drones in cluttered environments~\cite{busset2015acoustic}. 
Compared to these hardware-based solutions, RAN-side KPIs such as handover frequency and RSRP variance are attractive because they do not require new infrastructure. Our work builds on this idea, focusing specifically on the mobility and signal metrics available in the 5G RAN for detecting non-cooperative UAVs.

\subsection{UAVs in Border Security and Counter-UAS Systems}
\label{ssec:uavsOnBorderSecurity}

Border protection provides an especially relevant context for UAV intrusion detection. Koslowski and Schulzke documented how state agencies and nonstate actors deploy drones along borders, where state agencies use them for surveillance and the latter use them for smuggling or reconnaissance~\cite{koslowski_schulzke_drones_along_borders} to avoid state surveillance. Industry reports echo this trend as well. Airsight has detailed how adversaries exploit small drones for the delivery and surveillance of contraband in border zones~\cite{airsight_border_security_drones}. 
Traditional counter-UAS (C-UAS) systems in these settings combine radar, optical, and RF sensors, but rarely leverage cellular infrastructure. This creates a gap where 5G networks, already deployed in many border areas, could act not only as a communication infrastructure to be protected but also as part of the sense-detection-mitigation system cycle.

\subsection{LEO Satellite Backhaul and Fallback Mechanisms}
\label{ssec:satBackhual}

With the expansion of 5G into remote areas, low Earth orbit (LEO) satellite constellations are increasingly being used to provide backhaul connectivity. 
Huang et al. developed a multiagent deep reinforcement learning framework for a space–air–ground integrated network, in which UAV access and LEO satellite backhaul are jointly optimized. Their results show that throughput performance is sensitive to satellite handovers and resource allocation policies~\cite{huang2024sagin}. 
Cisco's industry evaluations confirm the viability of LEO-based 5G backhaul, reporting round trip latencies on the order of 50~ms, but also highlighting persistent jitter and occasional outages during satellite or gateway transitions~\cite{cisco_leo_wp_2025}. Research on NTN integration echoes these concerns. The case in which satellite backhaul is primary and a local fallback is needed for UAV intrusion mitigation has not been quantitatively studied. 
Our work addresses this gap by experimentally quantifying detect-to-mitigate delay under LEO outages and showing how local fallback deadlines cap vulnerability windows.

\subsection{Summary and Gap}
\label{ssec:summryGap}

Previous work has shown that UAVs exhibit anomalous mobility signatures in 5G networks~\cite{handoverchallenges, muzaffar2020first, mobilitystatedetection, 9319862}, that uncooperative UAVs can be detected using RF and RAN side metrics~\cite{abdalla2020uav, maksymiuk2022renyi, 8741970, busset2015acoustic}, and that drones are a growing challenge in border security~\cite{koslowski_schulzke_drones_along_borders, airsight_border_security_drones}. 
Meanwhile, the reliability limitations of the LEO backhaul are well established~\cite{huang2024sagin, cisco_leo_wp_2025}. However, no previous study has experimentally evaluated how these LEO backhaul impairments affect UAV intrusion response, nor measured the benefits of enforcing local fallback in this context. Our work contributes precisely to this missing piece: a quantitative analysis of fallback deadlines in mitigating rogue UAVs under LEO outage conditions.

\section{System Model and Assumptions}
\label{sec:sysModel+assupmtions}

\begin{figure}[t]
  \centering
  \includegraphics[width=\linewidth]{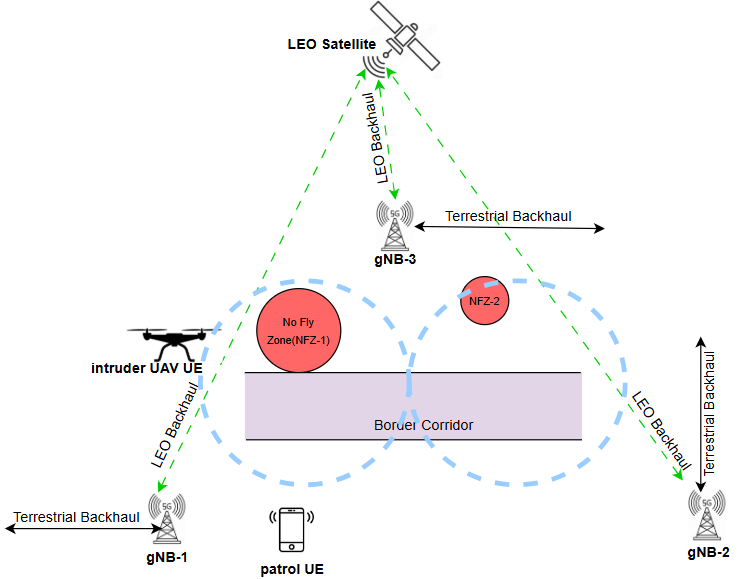}
  \caption{System model for UAV intrusion detection and mitigation in a border-security scenario}
  \label{fig:system model}
\end{figure}

We consider a simplified border-corridor topology illustrated as shown in Fig.~\ref{fig:system model} to show how UAV intrusion detection interacts with satellite backhaul impairments. The system shown in Fig.~\ref{fig:system model} includes a designated border corridor and nearby no-fly zones (NFZ) that represent sensitive or restricted airspace where UAV intrusion is not acceptable.

Wireless coverage is provided by multiple terrestrial gNBs positioned to ensure both corridor and NFZ coverage. This coverage is not for visual surveillance, but for cellular service. Authorized patrol UEs rely on it for normal connectivity, while intruder UAVs inevitably interact with gNBs in the zone when entering protected airspace. These interactions with the RAN form the basis for anomaly detection. For completeness of the illustration, Fig.~\ref{fig:system model} shows each gNB with terrestrial and satellite backhaul links. In practice, each gNB operates with a single backhaul type in a given scenario where some gNBs are configured with (idealized as reliable) terrestrial backhaul , while others are configured LEO satellite backhaul and thus subject to latency, jitter, and outages. This setup enables direct comparison between terrestrial-only control and satellite-impaired control.

Two types of user equipment (UEs) are modeled. The first are \textit{patrol UEs}, representing an authorized ground terminal that continuously moves along the boundaries of the corridor and acts as a baseline for normal ground mobility. Others are \textit{intruder UAVs},
representing an uncooperative aerial device. These UAVs first enter the corridor and then loiter within it using a periodic trajectory that ensures repeated exposure to different gNBs and NFZ regions. Patrol UEs are whitelisted as legitimate, whereas intruder UAVs with spoofed UEs are subject to detection and mitigation.

Propagation is modeled with free-space path loss and shadowing effects, while different penalties are applied to ground and UAV-bound UEs to capture antenna orientation and sidelobe reception. Additionally, when operating near any gNB, UAVs impose interference on patrol UEs, which is modeled as an extra downlink degradation before mitigation and partially
reduced after mitigation.

The handover logic follows a simplified version of 3GPP’s Event A3 rule: that is, a measurement-triggered condition in which a neighbor cell must become better than the serving cell by a configured offset (plus hysteresis) and stay that way for a duration known as Time-to-Trigger (TTT) before a handover is considered~\cite{etsi_ts_38_331_v15.02.01}. Once a UAV is detected, it can be locked down (pinned to its serving cell), preventing further handovers. Detection itself relies on observing mobility features such as frequent handovers or high variability in the received signal strength, provided that the UAV is not whitelisted. To ensure robustness, detection requires persistence over a short interval before triggering.

Mitigation commands are subject to backhaul conditions because, in our model, the lockdown decision is coordinated with a higher-level control entity (e.g., the core network or a remote command center). For purely terrestrial gNBs, this coordination is assumed to be immediate, so lockdown is applied as soon as detection occurs. However, for satellite-backhauled gNBs, the command must traverse the satellite link, introducing additional latency, jitter, and the possibility of outages that last for several seconds. Outages are modeled as stochastic processes that block commands for fixed durations. To protect against excessive delays, a local fallback mechanism can enforce a deadline, such as if a command is not successfully applied within a certain time (say, 2~s), the lockdown is triggered locally. Performance is evaluated using the following KPIs.
\begin{itemize}
    \item The delay between detection and mitigation,
    \item The number of UAV handovers during this delay window,
    \item The handover rate experienced by the ground patrol UE,
    \item The UAV’s dwell time in the corridor before being locked, and
    \item The number of NFZ violations.
\end{itemize}

Scenarios are defined as follows.
\begin{enumerate}
    \item \textbf{Terrestrial baseline:} all gNBs with terrestrial backhaul,
    \item \textbf{LEO-Outage:} satellite backhaul with outages but no fallback,
    \item \textbf{LEO-Outage+Fallback:} outage case with fallback enabled, and
    \item \textbf{Stress cases:} outage durations extended beyond fallback deadlines.
\end{enumerate}

In addition, a sensitivity analysis varies hysteresis, detection thresholds, and outage rates. Several simplifications are made to use control plane latency. For this, we simulate only one unrecognized UAV and one patrol UE, without traffic scheduling or HARQ. Antenna effects are modeled as simple penalties, and UAV interference is limited to degrading patrol UE's signal strength. These abstractions allow us to isolate the main research question of how satellite backhaul impairments and fallback deadlines affect the
detect-to-mitigate loop in UAV intrusion scenarios.

\section{Simulation Setup}
\label{sec:simulationSetup}

The experiments are carried out in a custom MATLAB-based simulator that implements the system model described in
Section~\ref{sec:sysModel+assupmtions}. The framework extends a minimal cellular control loop with modules for UAV's mobility, path loss and interference, handover logic, intrusion
detection, mitigation with satellite backhaul, and no-fly zone bookkeeping. The simulator is structured as a single MATLAB function with modular helpers, ensuring repeatability across scenarios and seeds.

\subsection{Parameter Grid and Seeds}
\label{ssec:parameterGrid+seeds}

The horizontal speed of the UAV intruder is varied between $speed\in\{6,9,12,15,18\}$ m / s, while the altitude is varied between $altitude\in\{60,90,120,150,180\}$ m. This results in $5\times 5=25$ (speed,altitude) grid points. For each point on the grid and scenario, $N=20$ independent Monte Carlo seeds are simulated. Randomness is present due to shadowing noise, outage realizations, and latency jitter.  The seeds of the random variables are deterministically derived from the $(\text{scenario}, speed, altitude, \text{rep})$ tuple to allow exact reproducibility.

Each simulation run spans $T_{\text{sim}}=200$ s with step size $\Delta t=0.5$ s, resulting in $400$ discrete time steps per run. This duration ensures that multiple UAV ingress and loiter cycles are captured, as well as several opportunities for detection and mitigation under backhaul outages.

\subsection{Detection and Handover Thresholds}
\label{ssec:detection}

Unless otherwise specified in the sensitivity runs, the handover hysteresis is set to $H=3$ dB with a time-to-trigger $\text{TTT}=1.5$ s. Intrusion detection operates with a sliding window of 20 s, requiring either at least three handovers in the window or a high variance in RSRP ($\geq 18$ dB$^2$) in the presence of at least three strong neighbors. Persistence of 3 s is required before flagging the UAV. Patrol UEs are exempt from detection using a whitelist.

\subsection{Backhaul Impairments}
\label{ssec:backhaulImpairments}

In our model gNBs with a satellite backhaul are subject to delay, jitter, and occasional outages. We set the mean latency to 30~ms with a jitter of 10~ms. These numbers are typical examples of the performance recently seen in tests of the low-earth orbit (LEO) satellite Internet. For example, Starlink backhaul field measurements report typical RTTs of 25–60~ms, with non-negligible variance due to routing and handovers~\cite{heine2023starlink, cisco_leo_wp_2025}. 

Outages are modeled according to a Poisson process with rates $\lambda\in\{0.01, 0.02, 0.05\}$Hz, corresponding to an event every 20–100 seconds on average. Each outage lasts 5~s in baseline runs, extended to 10~s in stress scenarios. These values are modeling assumptions intended to emulate the several-second disruptions that have been observed during satellite handovers and gateway transitions in real LEO constellations. For example, live streaming measurement studies on Starlink have recorded outages of the order of 1–2~s, with some application-layer rebuffering extending beyond 5~s under unfavorable conditions~\cite{fang2025robust, ma2022characteristics}. We chose 5-10 s outages longer than 1-2s seen in Starlink measurements because we wanted to stress-test the worse case stresses.

Finally, when the fallback is enabled, we enforce a local deadline $\tau=2$~s. If a lockdown command has not been applied by this time, the gNB executes it locally without waiting for backhaul recovery. This threshold is a design parameter chosen to represent a tight bound for security-sensitive scenarios, ensuring timely mitigation of intrusions. Although real deployments might configure different deadlines depending on risk tolerance and service requirements, the 2~s value serves to highlight the resilience gains achievable with local fallback.

\subsection{Key Performance Indicators}
\label{ssec:KPIs}

The following key performance indicators (KPIs) are collected per each run.
\begin{itemize}
    \item \textbf{Detect-Mitigate Delay:} elapsed time between UAV detection and the application of lockdown.
    \item \textbf{Extra Handovers:} number of transfer of UAVs in that gap, capturing instability during indecision.
    \item \textbf{Patrol UE HO Rate:} average handovers per minute for the patrolling UAVs/ UE, measuring collateral damage.
    \item \textbf{Dwell Time Before Lock:} time spent by the UAV in a border corridor after detection but before lockdown.
    \item \textbf{NFZ Violations:} number of discrete time steps in which the UAV entered a zone of no flying.
\end{itemize}

\subsection{Summary of Parameters}
\label{ssec:summaryOfParams}

Table~\ref{tab:params} summarizes the main simulation parameters.

\begin{table}[!htbp]
\centering
\caption{Simulation Parameters}
\label{tab:params}
\begin{tabular}{ll}
\hline
\textbf{Parameter} & \textbf{Value(s)} \\
\hline
Simulation time $T_{\text{sim}}$ & 200 s \\
Step size $\Delta t$ & 0.5 s \\
Carrier frequency $f_c$ & 3.5 GHz \\
Tx power & 46 dBm \\
Speeds $v$ & 6, 9, 12, 15, 18 m/s \\
Altitudes $z$ & 60, 90, 120, 150, 180 m \\
MC seeds $N$ & 20 per grid point \\
HO hysteresis $H$ & 3 dB (varied in sensitivity) \\
TTT & 1.5 s \\
Detection window & 20 s \\
HO threshold & 3 (varied in sensitivity) \\
RSRP variance threshold & 18 dB$^2$ \\
Strong neighbor threshold & 3 within $\Delta=6$ dB \\
Persistence & 3 s \\
Backhaul latency & 30 ms + jitter 10 ms \\
Outage rate $\lambda$ & 0.01, 0.02, 0.05 Hz \\
Outage duration & 5 s (10 s in stress) \\
Fallback deadline $\tau$ & 2 s \\
\hline
\end{tabular}
\end{table}

\section{Simulation Results}
\label{sec:SimResults}

This section presents the results of the simulation runs across the $(speed, altitude)$ grid, Monte Carlo seeds, and scenarios. Four families of figures are reported:

\begin{enumerate}
\item Detect-to-mitigate delay, 
\item Extra handovers incurred between detection and mitigation.
\item Collateral impact on ground patrol users
\item Stress scenarios under extended LEO outages. 
\end{enumerate}

These correspond to the key performance indicators (KPIs) described in Section~\ref{ssec:summaryOfParams}. Unless otherwise noted, plots show heat maps of median values between seeds or pooled empirical CDFs with bootstrap confidence intervals. Although NFZ violations were tracked as a KPI, in our runs they were rare and did not differ significantly between scenarios, so we omit detailed plots.

\subsection{Detect-to-Mitigate Delay}
\label{ssec:detectMitigateDelay}

\begin{figure}[!htbp]
\centering
\includegraphics[width=0.48\textwidth]{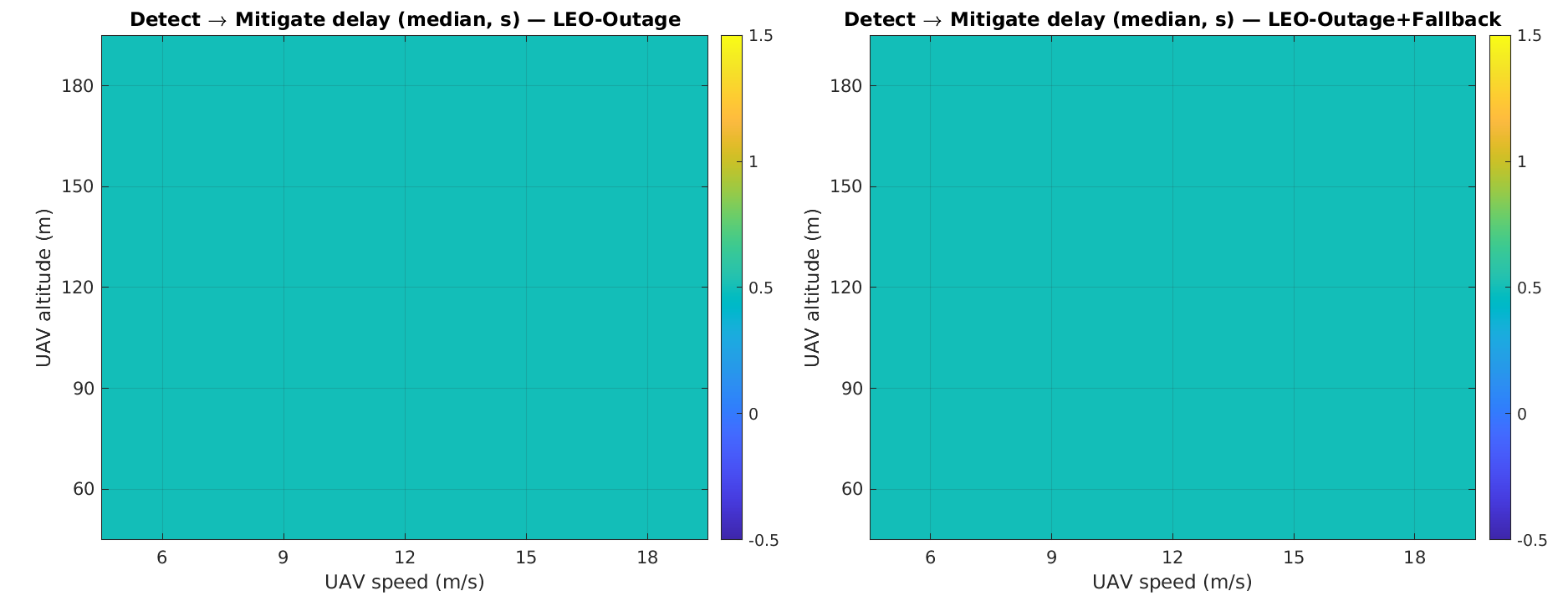}
\includegraphics[width=0.48\textwidth]{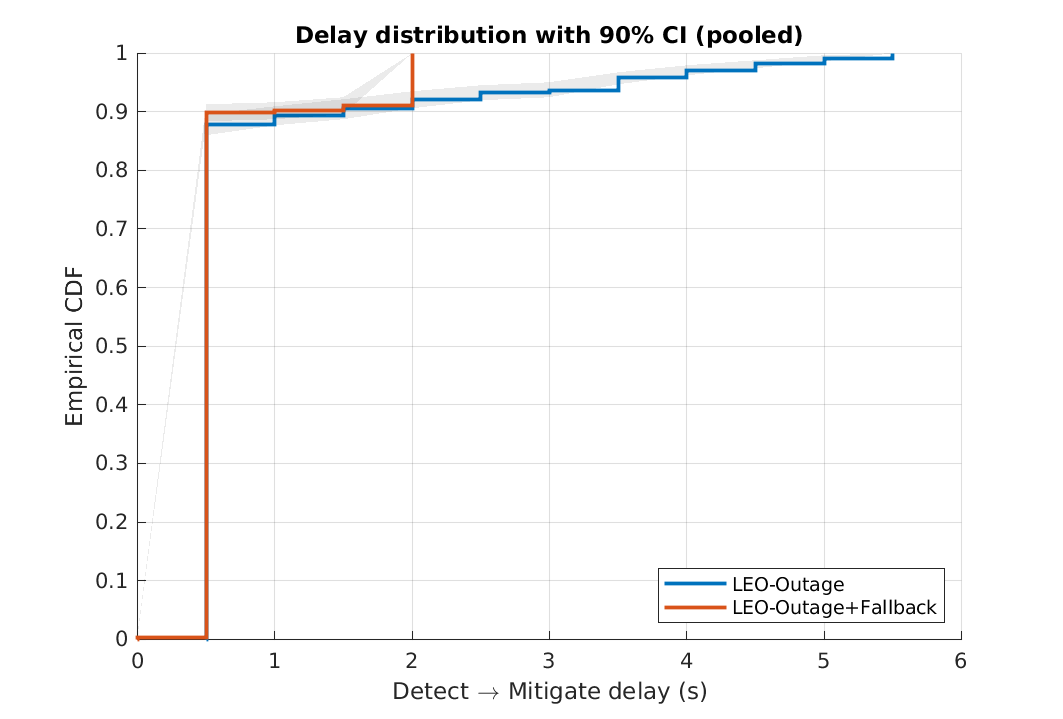}
\caption{Detect $\rightarrow$ Mitigate delay. Top: median heatmaps across the $(speed, altitude)$ grid. Bottom: pooled CDFs with bootstrap confidence bands.}
\label{fig:delay}
\end{figure}

Fig.~\ref{fig:delay} compares the elapsed time between UAV detection and lockdown application during LEO outages with and without fallback. In the pure LEO-Outage case, mitigation delay is dominated by the residual outage duration and backhaul latency jitter, with many runs exceeding several seconds. In contrast, enabling fallback sharply reduces the tail of the delay distribution, capping the worst-case delay to the 2~s deadline. The heat maps further confirm that this effect is
consistent across speeds and altitudes. while some variation exists due to outage realizations, the fallback scenario produces a much tighter delay band.

\subsection{Extra Handovers During Delay}
\label{ssec:exHandover}

\begin{figure}[!htbp]
\centering
\includegraphics[width=\linewidth]{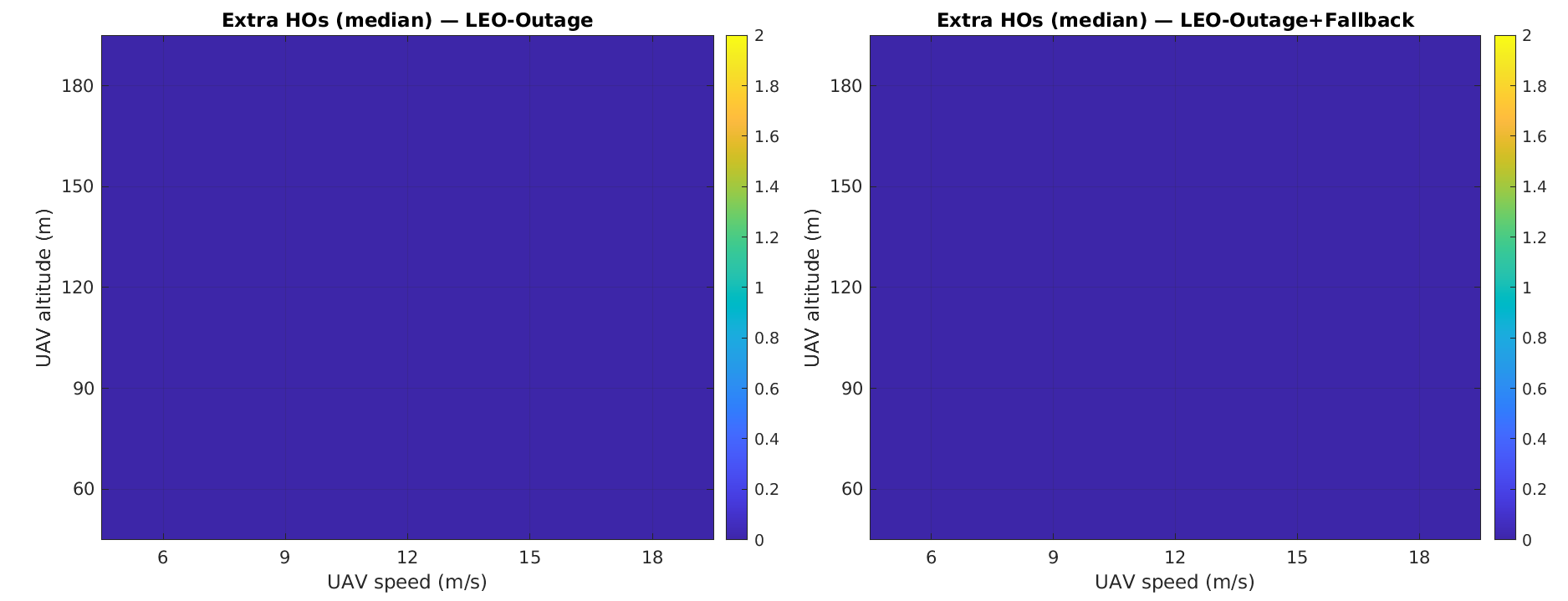}
\includegraphics[width=\linewidth]{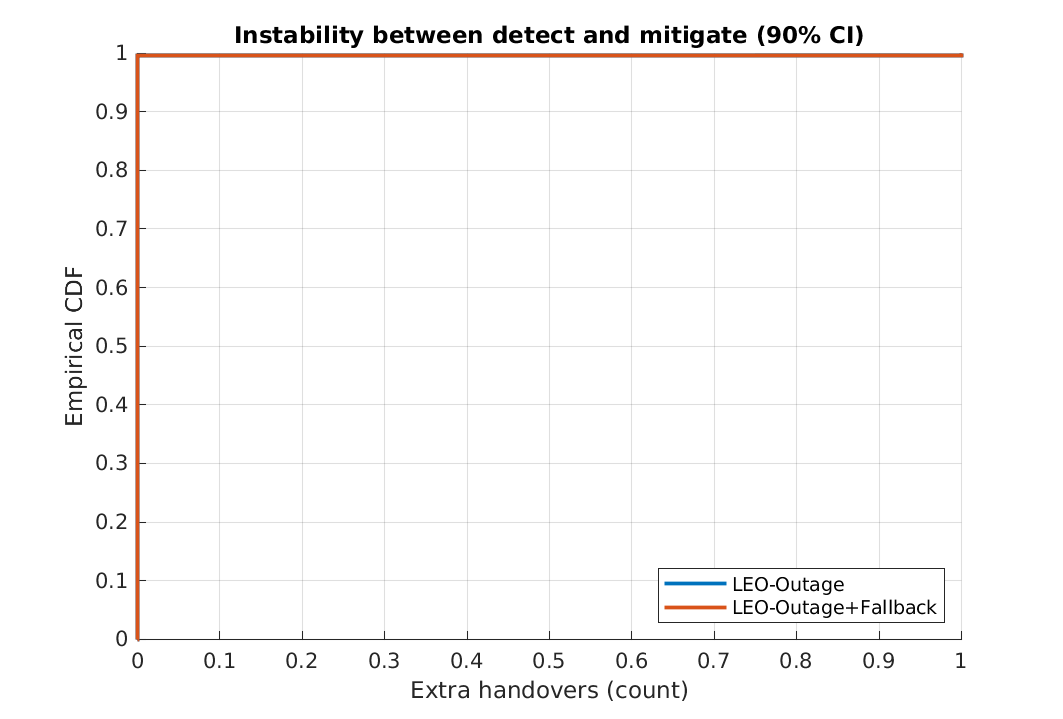}
\caption{Extra handovers between detection and mitigation. Top: median heatmaps. Bottom: pooled CDFs with confidence bands.}
\label{fig:extrahos}
\end{figure}

Fig.~\ref{fig:extrahos} shows the number of additional handovers incurred by the UAV between detection and mitigation. Across all scenarios, the distributions are nearly identical in both LEO-Outage and LEO-Outage + Fallback, and the vast majority of runs experience zero extra handovers. This indicates that, under our detection thresholds, mobility pattern and A3 settings, the UAV rarely has time to complete a handover in the short detection-to-mitigation window. Thus, the handover churn is not a dominant instability mechanism in this configuration.

\subsection{Collateral Impact on Ground Users}
\label{ssec:collateralImpact}

\begin{figure}[!htbp]
\centering
\includegraphics[width=\linewidth]{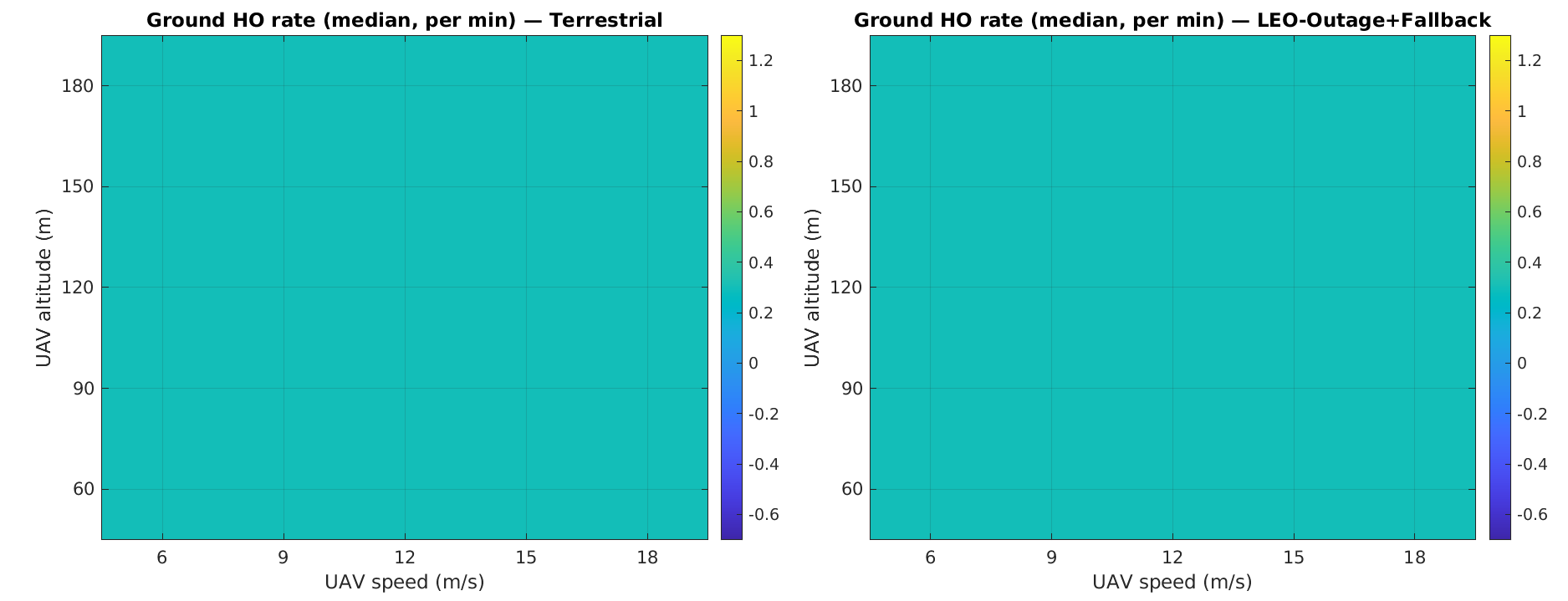}
\includegraphics[width=\linewidth]{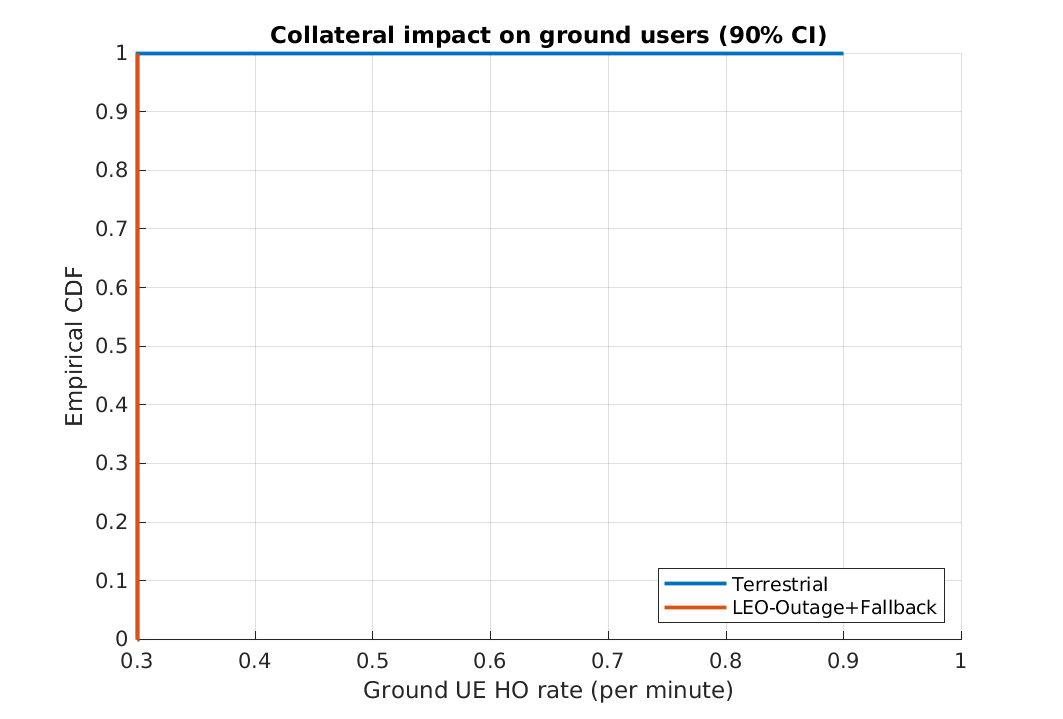}
\caption{patrol UE handover rate (per minute). Left: median heatmaps. Right: pooled CDFs.}
\label{fig:ground}
\end{figure}

Fig.~\ref{fig:ground} shows the average UE handover rate per minute for a baseline patrol UE. In both the terrestrial and LEO-Outage+Fallback scenarios, the distributions are essentially identical, with overlapping CDFs and uniform heatmaps. This indicates that the UAV lockdown mechanism does not introduce collateral instability for legitimate ground users. The results confirm that, under the conditions studied, fallback enforcement preserves terrestrial-level performance for authorized UEs.

\subsection{Stress Scenarios of Long Outages}
\label{ssec:longOutages}

\begin{figure}[!t]
\centering
\includegraphics[width=0.48\textwidth]{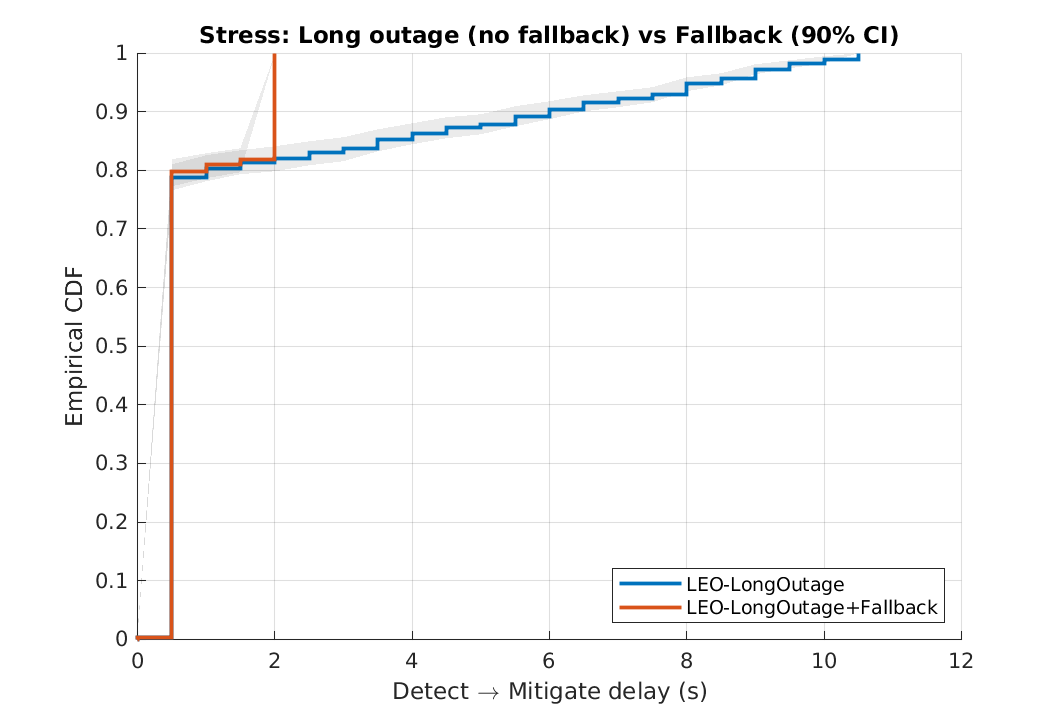}
\caption{Stress case: long outage (10~s) versus fallback with 2~s deadline. Fallback enforces a hard cap on mitigation delay.}
\label{fig:stress}
\end{figure}

Finally, Fig.~\ref{fig:stress} highlights a stress scenario in which LEO outages last significantly longer than the 2~s fallback deadline (e.g., 10~s). In the no-fallback case, mitigation delays extend linearly with the outage duration, resulting in unacceptably long vulnerable periods. With fallback, the delay is bounded at 2~s by local deadline enforcement, regardless of the duration of the outage. This scenario emphasizes that fallback is not merely an optimization, but an essential safeguard against rare but severe outages.

\subsection{Summary}
\label{ssec:summary}

Across all metrics, the addition of a 2 s local fallback mechanism provides a clear robustness gain that bounds mitigation delay and avoids collateral instability for patrol UEs. Although handover instability was hypothesized, our results show that extra handovers are negligible in both cases under the parameters studied. The results indicate that relying solely on satellite backhaul can be fragile in the face of outages, while hybrid designs with deadline-based fallback significantly improve resilience.

\subsection{Discussion and Implications}
\label{ssec:discussion}

The above results highlight three main insights that can be gained from our design and simulation of border-sensitive UAV intrusion detection and mitigation using 5G/6G networks.

\begin{enumerate}
\item \textbf{Importance of bounding mitigation delay}\\ 
Outage-prone satellite backhaul introduces stochastic delays that can exceed tolerable limits. The proposed fallback mechanism effectively caps these delays to a local deadline (2~s in this study), thus ensuring that mitigation actions occur within predictable limits. This is critical for security-sensitive applications, where excessive latency in lockdown could allow an intruder UAV to dwell inside a restricted corridor or no-fly zone.

\item \textbf{Reduction of instability and signaling churn}
We did not observe significant handover churn in our configuration, both with and without fallback, the UAV essentially did not incur additional handovers during the detect–mitigate gap. This suggests that with moderate hysteresis and TTT values, the system is relatively robust against short-term oscillations. Although the fallback mechanism did not materially reduce handover counts, its main benefit was the limitation delay in mitigation (as shown in Fig.~\ref{fig:delay}), which is the main vulnerability pathway in these scenarios.

\item \textbf{Minimal collateral harm to patrol UEs.}
A key concern with aggressive detection and lockdown is whether normal patrol UEs suffer from increased instability. The simulations show that ground HO rates remain close to the terrestrial baseline when fallback is enabled. This suggests that the mechanism can be deployed with a low
risk of penalizing legitimate users.

\end{enumerate}

\subsubsection{Implications for system design.}  
\label{sssec:implicationsSytemDesign}

The findings argue for a hybrid terrestrial-nonterrestrial
architecture where local control plane fallback complements satellite backhaul. Such designs balance coverage benefits from LEO satellites with the robustness of terrestrial enforcement. Furthermore, stress scenarios demonstrate that fallback is not only a performance optimization but a necessary safeguard against rare and catastrophic outages. Future 6G standards and deployments should explicitly account for such hybrid resilience mechanisms.

\subsubsection{Limitations}  
\label{sssec:limitations}

The current study models a simplified border corridor with a single intruder UAV and a single patrol UE, all using 5G/6G connectivity. Our model does not capture full interference coupling, mobility diversity, or large-scale multicell coordination. The negligible extra handovers observed may be due to current hysteresis and TTT settings, and different parameters could make the handover churn more visible. Thus, while trends are robust, absolute KPI values should be interpreted with caution. Expanding to multi-UAV scenarios and more complex backhaul dynamics remains important ongoing future work.

\section{Conclusions}
\label{sec:conclusions}

This paper presented a minimal end-to-end simulator for
evaluating intrusion detection and mitigation of uncooperative UAVs in a secure land border. The system model integrates terrestrial gNBs, LEO satellite backhaul with stochastic outages, and a detection logic based on handover dynamics and RSRP variance. A simple lockdown mitigation mechanism was applied once an intruder was detected.

Using Monte Carlo sweeps of UAV speeds, altitudes,
and backhaul scenarios, the study yielded several insights.
First, satellite outages can cause unbounded delays in mitigation, but a local fallback deadline effectively caps these delays, providing predictable response times even under stochastic impairments. Second, the collateral impact on patrol UEs remains minimal, with handover rates close to terrestrial baselines. Third, while handover instability was hypothesized as a potential issue, our results show that under the studied
detection thresholds and mobility patterns, the intruder UAV incurred essentially no additional handovers in the detect-mitigate window. This finding is itself instructive, as it suggests that, for moderate A3 parameters of 3GPP, delay bounding emerges as the dominant resilience lever under our chosen parameters, while handover churn may only emerge
under more aggressive thresholds or higher mobility.

Stress scenarios caused by longer delays of backhaul outages on satellite links further showed that fallback is indispensable to prevent extreme cases of vulnerability induced by long outages. These findings support the design of hybrid terrestrial-nonterrestrial architectures in which local control complements satellite coverage, ensuring robust and predictable
responses to UAV intrusions.

Our ongoing work is extending this framework to multi-UAV scenarios, UAVs that do not call home for navigation with more realistic mobility traces, and richer interfeence models. Ultimately, in addition to supporting the primary goal of supporting border protection, a secondary goal is to provide actionable information on standards and deployments that must protect borders against uncooperative aerial threats while protecting legitimate terrestrial users.

\bibliographystyle{IEEEtran}
\bibliography{refs}

\end{document}